\documentclass[]{aa} 

\usepackage[dvipsnames]{xcolor}

\usepackage{tabularx}
\usepackage{array}

\newcolumntype{Y}{>{\centering\arraybackslash}X}

\usepackage{graphicx}
\usepackage{txfonts}
\usepackage{hyperref}

\begin{document} 

\newcommand{\afffias}{Frankfurt Institute for Advanced Studies (FIAS), Ruth-Moufang-Strasse~1, 60438 Frankfurt am Main, Germany}
\newcommand{\affbgu}{Physics Department, Ben-Gurion University of the Negev, Beer-Sheva 84105, Israel}
\newcommand{\affbul}{Institute for Nuclear Research and Nuclear Energy, Bulgarian Academy of Sciences, Sofia, Bulgaria}

\title{Late-time constraints on interacting dark energy: Analysis independent of $H_0$, $r_d$, and $M_B$}

\author{David Benisty\inst{1,2} \and  Supriya Pan\inst{3,4} \and Denitsa Staicova\inst{5} Eleonora Di Valentino \inst{6} Rafael C. Nunes \inst{7,8}}

\institute{$^{1}\,$ Frankfurt Institute for Advanced Studies (FIAS), Ruth-Moufang-Strasse~1, 60438 Frankfurt am Main, Germany 
\\$^{2}\,$ Helsinki Institute of Physics, P.O. Box 64, FI-00014 University of Helsinki, Finland
\\$^{3}\,$ Department of Mathematics, Presidency University, 86/1 College Street, Kolkata 700073, India
\\$^{4}\,$ Institute of Systems Science, Durban University of Technology, PO Box 1334, Durban 4000, Republic of South Africa
\\$^{5}\,$ Institute for Nuclear Research and Nuclear Energy, Bulgarian Academy of Sciences, Sofia, Bulgaria
\\$^{6}\,$ School of Mathematics and Statistics, University of Sheffield, Hounsfield Road, Sheffield S3 7RH, United Kingdom
\\$^{7}\,$ Instituto de F\'{i}sica, Universidade Federal do Rio Grande do Sul, 91501-970 Porto Alegre RS, Brazil
\\$^{8}\,$ Divis\~{a}o de Astrof\'{i}sica, Instituto Nacional de Pesquisas Espaciais, Avenida dos Astronautas 1758, S\~{a}o Jos\'{e} dos Campos, 12227-010, S\~{a}o Paulo, Brazil}
   \date{}

 
\abstract{We investigated a possible interaction between cold dark matter and dark energy, corresponding to a well-known interacting dark energy model discussed in the literature within the context of resolving the Hubble tension. We put constraints on it in a novel way, by creating new likelihoods with an analytical marginalization over the Hubble parameter $H_0$, the sound horizon $r_d$, and the supernova absolute magnitude $M_B$.  Our aim is to investigate the impacts on the coupling parameter of the interacting model, $\xi$, and the equation of state of dark energy $w$ and the matter density parameter $\Omega_{m,0}$. The late-time cosmological probes used in our analysis include the PantheonPlus (calibrated and uncalibrated), cosmic chronometers, and baryon acoustic oscillation samples and the Pantheon for comparison. Through various combinations of these datasets, we demonstrate hints of an up to $2\sigma$ deviation from the standard $\Lambda$ cold dark matter model.}
  


%

\maketitle
\section{Introduction}
\label{sec:Introduction}
Over the last few decades, cosmological measurements indicating  an acceleration in the expansion of the Universe   have  suggested that Einstein's general theory of relativity (GR) alone is probably not the ultimate theory of gravity capable of explaining all the available observational  evidence. Observational data from Type Ia supernovae (SNeIa)~\cite{SupernovaSearchTeam:1998fmf,SupernovaCosmologyProject:1998vns,Pan-STARRS1:2017jku}, baryon acoustic oscillations (BAOs)~\cite{Addison:2013haa,Aubourg:2014yra,Cuesta:2014asa,Cuceu:2019for}, and the cosmic microwave background (CMB)~\cite{Planck:2018vyg} provide compelling evidence for modifications either in the matter sector of the Universe or in the gravitational sector. The simplest modification is the introduction of a positive cosmological constant, $\Lambda$, into the gravitational equations described by Einstein's GR~\cite{SupernovaSearchTeam:1998fmf,SupernovaCosmologyProject:1998vns,Weinberg:1988cp,Lombriser:2019jia,Copeland:2006wr,Frieman:2008sn} and the resulting picture ---the so-called $\Lambda$-cold dark matter ($\Lambda$CDM) cosmological model --- has been found to be consistent with a wide range of observational datasets. Nevertheless, the $\Lambda$CDM model is now facing both theoretical and observational challenges~\citep{Verde:2019ivm,Riess:2019qba,DiValentino:2020zio,Riess:2021jrx,DiValentino:2020vvd,DiValentino:2020srs}. Consequently, there is growing demand for a revision of $\Lambda$CDM cosmology~\citep{Knox:2019rjx,Jedamzik:2020zmd,DiValentino:2021izs,Abdalla:2022yfr,Kamionkowski:2022pkx,Escudero:2022rbq,Vagnozzi:2023nrq,Khalife:2023qbu}. Thus, the question arises as to whether GR$\,+\,\Lambda$ is the fundamental theory of gravity, or merely an approximation of a more complete gravitational theory yet to be discovered. One natural avenue of exploration is to consider modified gravity theories, which show theoretical and observational promise in addressing the observed discrepancies. With the ever-increasing sensitivity and precision of present and upcoming astronomical surveys, modified gravity theories emerge as viable contenders alongside GR$\, +\, \Lambda$. The search for the ultimate  answer in this direction is ongoing. According to the existing literature, we currently have a cluster of cosmological scenarios broadly classified into two categories: (i) cosmological scenarios within GR, commonly known as dark energy models, and (ii) cosmological scenarios beyond GR, commonly known as modified gravity models.

In this article, we focus on the first approach, which means that the gravitational sector of the Universe is well described by GR, but modifications of the matter fields are needed to explain the current accelerating phase and recent observational tensions and anomalies that persist in the structure of the standard cosmological model. The list of cosmological models in this particular domain is extensive, and here we are interested in investigating one of the generalized and more appealing cosmological theories in which dark matter (DM) and dark energy (DE) interact with each other via an energy exchange mechanism between them. The theory of interacting DM--DE, widely known as IDE, has garnered significant attention in the community and has been extensively studied, with many promising results~\cite{Amendola:1999er,Cai:2004dk,Barrow:2006hia,Valiviita:2008iv,Valiviita:2009nu,Gavela:2010tm,Chen:2011cy,Clemson:2011an,Salvatelli:2013wra,Li:2013bya,Salvatelli:2014zta,Yang:2014gza,Yang:2014okp,Faraoni:2014vra,Yang:2014hea,Li:2014cee,Pan:2012ki,Nunes:2016dlj,Yang:2016evp,DiValentino:2017iww,Mifsud:2017fsy,Yang:2017zjs,Yang:2017ccc,Pan:2017ent,Yang:2018uae,DiValentino:2019ffd,Yang:2018pej,Yang:2018xlt,Yang:2018ubt,Yang:2018uae,DiValentino:2019jae,Pan:2019gop,Pan:2020zza,DiValentino:2020kpf,Johnson:2020gzn,Johnson:2021wou,Gao:2021xnk,Escamilla:2023shf,Zhai:2023yny,Hoerning:2023hks,Pan:2023mie,Silva:2024ift,Giare:2024smz} (also see~\cite{Bolotin:2013jpa,Wang:2016lxa,Wang:2024vmw}).  The IDE models gained prominence in modern cosmology due to their capacity to alleviate tensions in some key cosmological parameters (see ~\cite{DiValentino:2017iww,Yang:2018euj,DiValentino:2019ffd,Aloni:2021eaq,Khachatryan:2020reb,Liu:2022hpz,Wagner:2022etu,Zhao:2022ycr,Vagnozzi:2023nrq,Pan:2023mie,Wang:2024vmw} for those alleviating the Hubble constant tension, and see~\cite{Pourtsidou:2016ico,An:2017crg,Benisty:2020kdt,Nunes:2021ipq,Lucca:2021dxo,Joseph:2022jsf,Naidoo:2022rda} for  those alleviating the growth tension). In IDE, the coupling function (also known as the interaction function) characterizing the transfer of energy between the dark components is the only ingredient that plays an effective role, and its non-null value indicates a deviation from the $\Lambda$CDM cosmology. The coupling between the dark sectors therefore invites new physics beyond the $\Lambda$CDM paradigm and could offer interesting possibilities in cosmology.

In IDE, the choice of the coupling function is not unique, and there is  freedom to explore a variety of interaction functions with the available observational data. In the current work, we investigate a particular well-known interacting model~\cite{Wang:2016lxa,DiValentino:2017iww,
DiValentino:2019ffd,Yang:2019uog}  using the latest available data from SNeIa, transversal BAO measurements, and the cosmic chronometers (CCs). We adopt a model-independent approach to address the cosmological tensions. Instead of assuming any prior knowledge about specific model parameters related to these tensions in cosmology, we choose to marginalize over these parameters. This means, we integrate out these parameters from the resulting $\chi^2$, ensuring that our results become independent of them.

The marginalization approach, which stems from previous works~\cite{DiPietro:2002cz,Nesseris:2004wj,Perivolaropoulos:2004yr,Lazkoz:2005sp,Basilakos:2016nyg,Anagnostopoulos:2017iao,Camarena:2021jlr}, was recently used by~\cite{Staicova:2021ntm} to study DE models by marginalizing over $H_0 \cdot r_d$.~\cite{Staicova:2021ntm}  demonstrated that the preference for a DE model can be highly sensitive to the choice of the BAO dataset when using the marginalization approach. For this reason, we adopt the same methodology here but in the context of an interaction scenario between DM and DE. Our goal is to clarify the robustness of previous results when the tension parameter is marginalized and to assess the sensitivity to the choice of the SNeIa dataset

The paper is organized as follows. In Section \ref{sec:model}, we describe the gravitational equations of an IDE scenario and then we propose two different IDE models distinguished by the equation of state of the DE sector 
that we study in this work. In Section \ref{sec:method}, we describe the methodology and the observational datasets used to constrain these interacting cosmological scenarios. Then, in Section \ref{sec-results}, we present the results of the interacting scenarios proposed in this work. Finally, in Section \ref{sec-conclusions}, we present our main conclusions.

\section{Interacting dark matter and dark energy}
\label{sec:model}

We work under the assumption of a spatially flat Friedmann-Lema\^{i}tre-Robertson-Walker (FLRW) line element: 
\begin{equation}
ds^2 = -dt^2 + a^2(t) \left[dr^2 + r^2 (d\theta^2 + \sin^2 \theta d\phi^2)\right],
\end{equation}
where $a(t)$ is the scale factor of the Universe. 
We consider that the matter sector of the Universe is minimally coupled to gravity as described by Einstein's GR and the matter sector is comprised of nonrelativistic baryons and two perfect dark fluids, namely pressureless DM and DE. 
In the presence of a nongravitational  interaction between DM and DE, which is characterized by a coupling function $Q (t)$, also known as the interaction rate, 
the continuity equations of the dark fluids can be written as~\cite{Wang:2016lxa}:

\begin{eqnarray}\label{dark}
\left\{\begin{array}{ccc}
\dot{\rho}_{\rm CDM} + 3 H \rho_{\rm CDM} &=&-Q (t)\\
\bigskip \\
\dot{\rho}_{\rm DE} + 3 H (1+w) \rho_{\rm DE} &=& Q(t), 
\end{array}\right.
\end{eqnarray}
where an overhead dot represents the derivative with respect to the cosmic time; $\rho_{\rm CDM}$, $\rho_{\rm DE}$ are the energy density for pressureless DM and DE, respectively; $w$ represents the barotropic equation of state for the DE fluid; and $H$ is the Hubble parameter. The Hubble parameter connects the energy densities of the matter sector as
\begin{eqnarray}\label{Hubble}
\rho_{\rm b} + \rho_{\rm CDM} + \rho_{\rm DE} = \left(\frac{3}{8 \pi G}\right) H^2,
\end{eqnarray}
where  $G$ is Newton's gravitational constant and $\rho_{\rm b}$ denotes the energy density of baryons, and it follows the usual evolution law $\rho_{b} \propto (a/a_0)^{-3}$, in which $a_0$ is the scale factor at the present time, set to unity. As we are only interested in the background dynamics at late times, we neglect the radiation contribution. Finally, $Q (t)$ in Eq.~(\ref{dark}) denotes the interaction function indicating the rate of energy transfer between the dark components. We note that $Q (t) > 0$ indicates the transfer of energy from DM to DE and $Q (t) < 0$ indicates that energy flow takes place from DE to DM. Once the interaction function is prescribed, the background dynamics of the model can be determined using the conservation equations (\ref{dark}),  together with the Friedmann equation (\ref{Hubble}).  In this article, we focus on the spatially flat case. 

Now,  there are precisely two approaches to selecting an interaction function. One can either derive this interaction function from some fundamental physical theory, or, alternatively, consider a phenomenological choice of the interaction function and test it using observational data. Although the former approach is theoretically robust and appealing, the quest for a difinitive solution in this regard is still ongoing. 
We consider a well-known interaction function of the form~\cite{Wang:2016lxa}:
\begin{eqnarray}\label{interaction-function}
 Q = 3 H \xi \rho_{\rm DE},    
\end{eqnarray}
which was initially motivated on phenomenological grounds, but is not robustly backed up theoretically~\cite{Pan:2020zza}. In the expression for $Q$ in Eq.~(\ref{interaction-function}), $\xi$ refers to the coupling parameter of the interaction function and in general this could be either constant or time dependent \cite{Chen:2011cy,Faraoni:2014vra,Yang:2019uzo}. Although the time-varying $\xi$ is expected to offer a more generalized interacting dynamics, in this work we consider that $\xi$ is constant.  
We note that for some coupling functions, the energy density of one or both of the dark fluids could be negative \cite{Pan:2020mst}. However, we do not impose any further constraints on the proposed model and allow the observational data to decide the fate of the resulting cosmological scenario. As  at present the community is expressing an interest in the negativity of the energy density of the DE or the cosmological constant~\cite{Poulin:2018zxs,Wang:2018fng,Visinelli:2019qqu,Calderon:2020hoc}, we keep this issue open for the future.

In this work, we explore two distinct IDE 
scenarios:
{\it (i)} the IDE 
scenario where the EoS of DE takes the simplest value of $w = -1$, denoted as the ``$\xi$CDM'' model and {\it (ii)} the IDE 
scenario where $w \neq -1$ is constant  but is a parameter that is free to vary in a certain region,  referred to as the  ``$w\xi$CDM'' model.

Given the coupling function (\ref{interaction-function}), and assuming our general case with the DE equation of state, $w$, the evolution laws of DE and CDM can be analytically obtained as  

\begin{eqnarray}\label{solutions}
\left\{\begin{array}{ccc}
\rho_{\rm DE} =  \rho_{\rm DE, 0}\; a^{-3 (1+w - \xi)},\\
\bigskip\\
\rho_{\rm CDM} = \rho_{\rm CDM,0}a^{-3} + \frac{\xi \rho_{\rm DE,0}}{\xi -w} \bigg( a^{-3} - a^{-3 (1+w-\xi)}\bigg). \end{array}\right.  
\end{eqnarray}
Consequently, the dimensionless Hubble parameter can be expressed as   \
\begin{eqnarray}
E(z)^2 = \Omega_{\rm m,0} (1+z)^3 + \Omega_{\rm DE,0} (1+z)^{3 (1+w -\xi)}  \nonumber\\+ \frac{\xi \Omega_{\rm DE,0}}{\xi - w} \bigg((1+z)^3 - (1+z)^{3 (1+w -\xi)} \bigg), 
\label{Eqs:E2}
\end{eqnarray}
where $1 + z = a_0/a$, and $\Omega_i = \rho_{i}/\rho_{c}$ ($i = {\rm CDM}, {\rm DE}$) is the density parameter of the $i$-th fluid (consequently, $\Omega_{i, 0}$ represents the current value of the same parameter), where $\rho_c = 3H^2/8\pi G$ is the critical density of the Universe. We note that 
$\Omega_{\rm m,0} = \Omega_{\rm CDM,0}+\Omega_{\rm b,0}$. From the initial condition $E (0) = 1,$ we get $\Omega_{\rm m,0} + \Omega_{\rm DE,0} = 1$.

\section{Methodology}
\label{sec:method}

In this section, we describe the marginalization procedure that has been adopted in this article in an attempt to constrain the proposed cosmological interaction scenarios. 

\subsection{Marginalization over degenerate parameters}
To circumvent the Hubble tension and mitigate the degeneracy between $H_0$ and $r_d$ in BAO data, we redefine the $\chi^2$ to integrate variables we prefer not to directly handle. In a cosmological model with $n$ free parameters (e.g., $\Omega_{m,0}$, $\xi$, $w$, etc.), these parameters are constrained by minimizing the $\chi^2$ function:
\begin{equation}
\begin{split}
\chi^2 = \sum_{i} \left[\vec{v}_{obs} - \vec{v}_{model}\right]^{T} C_{ij}^{-1} \left[\vec{v}_{obs} -  \vec{v}_{model} \right], 
\end{split}
\end{equation} 
where $\vec{v}_{\text{obs}}$ represents the vector of observed points at each $z$, $\vec{v}_{\text{model}}$ denotes the theoretical prediction of the model, and $C_{ij}$ is the covariance matrix. For uncorrelated data, $C_{ij}$ reduces to a diagonal matrix with the errors ($\sigma_i^{-2}$) on the diagonal. 

In our analysis, we use three distinct datasets: the SNeIa datasets (Pantheon, calibrated, and uncalibrated PantheonPlus), the transverse BAO dataset, and the CC dataset. Below, we delineate the marginalization process for each of these datasets.

\subsection{BAO redefinition}

For the BAO dataset, we use the angular scale measurement $\theta_{\text{BAO}}(z)$, which provides the angular diameter distance $D_A$ at redshift $z$:
\begin{equation}
\theta_{BAO}\left(z\right) = \frac{r_d}{\left(1+z\right)D_A(z)} = \frac{H_0 r_d}{c} h(z),
\end{equation}
where
{\begin{equation}
h\left(z\right) = \frac{1}{ (1+z) f(z)}
\end{equation}}
and
\begin{equation}
f\left(z\right) = \frac{1}{(1+z)  }\int \frac{dz'}{E(z')}. 
\end{equation}
This is valid for the flat universe case.
We can express the vector as a dimensionless function multiplied by the parameter $c/H_0 r_d$:
\begin{equation}
\vec{v}_{model} = \beta \left(f(z) , E(z)^{-1}\right) = \beta \vec{f}_{model}.
\end{equation}

By following the approach in~\cite{Lazkoz:2005sp,Basilakos:2016nyg,Anagnostopoulos:2017iao,Camarena:2021jlr,Staicova:2021ntm}, one can isolate $\frac{c}{H_0 r_d}$ in the $\chi^2$ expression by expressing it as
\begin{equation}
\chi^2 = \beta^2 A_{\theta} - 2 B_{\theta} \beta + C_{\theta},
\end{equation} 
where $\beta=\frac{H_0 r_d}{c}$ and 
\begin{subequations}
\begin{equation}
A_{\theta}  = \sum_{i = 1}^{N} \frac{h(z_i)^2}{\sigma_i^2},
\end{equation}
\begin{equation}
B_{\theta}  = \sum_{i = 1}^{N} \frac{{\theta}_{D}^{i}\, h(z_i) }{\sigma_i^2},
\end{equation}
\begin{equation}
C_{\theta}  = \sum_{i = 1}^{N} \frac{ \left(\theta_{D}^{i}\right)^2}{\sigma_i^2}.
\end{equation}
\end{subequations}

Using Bayes's theorem and marginalizing over $\beta=H_0 r_d/c$, we obtain
\begin{equation}
p\left(D,M\right) = \frac{1}{p\left(D|M\right)} \int \exp\left[-\frac{1}{2}\chi^{2}\right] d\beta,
\end{equation}
where $D$ represents the data used and $M$ denotes the model. Consequently, by employing $\widetilde{\chi}^2_{BAO} = -2 \ln p\left(D,M\right)$, we derive the marginalized $\chi^2$ in the form:
\begin{equation}
\widetilde{\chi}^2 = C_\theta-\frac{B_\theta^2}{A_\theta} + \log\left(\frac{A_\theta}{2 \pi}\right).
\label{eq:chi2BAO}
\end{equation} 
This $\widetilde{\chi}^2_{\theta}$ depends solely on $h(z)$, with no dependence on $H_0 \cdot r_d/c$.

\subsection{Supernova redefinition}
Similarly, by following the approach used in~\cite{DiPietro:2002cz,Nesseris:2004wj,Perivolaropoulos:2004yr,Lazkoz:2005sp,Benisty:2023laj}, we integrate over $M_B$ and $H_0$ to derive the integrated $\chi^2$. The measurements of SNeIa are described by the luminosity distance $d_L(z)$ (related to $D_A$ by $D_A=d_L(z)/(1+z)^2$) and its distance modulus $\mu(z)$, which is given by

\begin{equation}
        \mu_B (z) - M_B = 5 \log_{10} \left[ d_L(z)\right] + 25  \,,
\label{eq:dist_mod_def}
\end{equation}
where $d_L$ is measured in megaparsecs (Mpc), and $M_B$ represents the absolute magnitude. 

For these, one can obtain the following integrated ${\chi}^2_{SN}$:
\begin{equation}
\widetilde{\chi}^2_{SN} = D-\frac{E^2}{F} + \ln\frac{F}{2\pi},
\end{equation}
where
\begin{subequations}
\begin{equation}
D = \sum_i \left( \Delta\mu \, C^{-1}_{\rm cov} \, \Delta\mu^T \right)^2,
\end{equation}
\begin{equation}
E = \sum_i \left( \Delta\mu \, C^{-1}_{\rm cov} \, \boldsymbol{\mathit{E}} \right),
\end{equation}
\begin{equation}
F = \sum_i  C^{-1}_{\rm cov}  ,
\end{equation}
\end{subequations}
where $\Delta\mu =\mu_{}^{i} - 5 \log_{10}\left[d_L(z_i)\right]$, $\boldsymbol{\mathit{E}}$ represents the unit matrix, and $C^{-1}_{cov}$ is the inverse covariance matrix of the dataset. For the Pantheon dataset, the total covariance matrix is given by $C_{\rm cov}=D_{\rm stat}+C_{\rm sys}$, where $D_{\rm stat}=\sigma_i^2$ arises from the measurement and $C_{\rm sys}$ is provided separately~\cite{Deng:2018jrp}. For PantheonPlus, the covariance matrix already includes both the statistical and systematic errors.

\subsection{Cosmic chronometers redefinition}
Following the same procedure as described in~\cite{Camarena:2021jlr}, but for the CC likelihood, $\chi^2_{CC}$, we obtain
\begin{equation}
\chi^2_{CC}=\frac{(H_0 E(z)-H_{obs}(z))^2}{\sigma^2}
.\end{equation} 

When applied to correlated data with a covariance matrix, the $\chi^2$ expression is redefined as
\begin{equation}
\chi^2_{CC}=-\left(G - \frac{B^2}{A} + log\left(\frac{A}{2\pi}\right)\right)
,\end{equation}
where
\begin{subequations}
\begin{equation}
G = \sum_i \left( H_i \, C^{-1}_{cov} \, H_i^T \right),
\end{equation}
\begin{equation}
B = \sum_i \left( E_i \, C^{-1}_{cov} \,  H_i \right),
\end{equation}
\begin{equation}
A = \sum_i (E_i C^{-1}_{cov}  E_i^T),
\end{equation}
\end{subequations}
where $H_i= H (z_i, \theta)$ represents the observational data points at each $z$, and $E_i= E(z_i, \theta)$ denotes the theoretical predictions for $E(z)$.

\subsection{Combined analysis}

In our analysis, we also consider the combined likelihood as follows:
\begin{equation}
\widetilde{\chi}^{2} = \widetilde{\chi}_{BAO}^{2} + \widetilde{\chi}_{SN}^{2} + \widetilde{\chi}_{CC}^{2}.
\end{equation}
The $\widetilde{\chi}^{2}$ depends only on the total energy density and the interaction strength. It is important to note that the above $\chi^2$ is not normalized, meaning that its absolute value does not serve as a useful measure of the quality of the fit.

\subsection{Datasets and priors}
\label{subsec-data}

In this work, we consider the following datasets:

\paragraph{BAO:} For BAO, we adopt the transversal angular dataset provided by~\cite{Nunes:2020hzy}. These points exhibit minimal dependence on the cosmological model, rendering them suitable for testing various DE models. While they are uncorrelated, the methodology's minimal assumptions on cosmology result in larger errors compared to those obtained using the standard fiducial cosmology approach~\cite{Bernui:2023byc,Nunes:2020uex}.

\paragraph{SNeIa:} For the SNeIa dataset, we use three different compilations, as described below:

\begin{itemize}
\item PantheonPlus and SH0ES (labeled PP): The PantheonPlus dataset, along with its covariance, comprises 1701 light curves of 1550 spectroscopically confirmed SNeIa, from which distance modulus measurements have been derived~\cite{Riess:2021jrx, Brout:2022vxf, Scolnic:2021amr}. Compiled across 18 different surveys, these light curves represent a significant enhancement over the initial Pantheon analysis, particularly at low redshifts.\footnote{ \url{https://github.com/PantheonPlusSH0ES/DataRelease}}

\item PantheonPlus
with removed SH0ES calibration (labeled PPNoS): 
The PP dataset includes the SH0ES light curves for SNeIa with $z<0.01$ along with their combined systematic covariance. To exclusively utilize the PantheonPlus dataset, we excluded all objects with $z<0.01$ and removed their covariance from the overall covariance matrix.

\item Pantheon (labeled as P): For comparison purposes, we include the ``old'' Pantheon dataset along with its covariance matrix. This dataset comprises $1048$ SNeIa luminosity measurements in the redshift range $z\in (0.01,2.3)$, binned into 40 points~\cite{Pan-STARRS1:2017jku}. Additionally, we incorporate systematic errors provided by the binned covariance matrix.\footnote{\url{https://github.com/dscolnic/Pantheon/}}
\end{itemize}

\paragraph{CC:} The CC dataset is based on the differential ages of passive galaxies (cosmic chronometers)~\cite{Moresco:2012jh,  Moresco:2015cya, Moresco:2016mzx}. We use the most recent version of the CC dataset, which includes the full covariance matrix accounting for systematic uncertainties stemming from the initial mass function, stellar library, and metallicity, which has been published in~\cite{Moresco:2020fbm}.\footnote{\url{https://gitlab.com/mmoresco/CCcovariance}}

For likelihood maximization, we employed an affine-invariant nested sampler, as implemented in the open-source package \texttt{Polychord}~\cite{Handley:2015fda}, and the results are presented using the \texttt{GetDist} package~\cite{Lewis:2019xzd}. Convergence in \texttt{Polychord} is achieved when the posterior mass contained in the live points reaches $p=10^{-2}$ of the total calculated evidence. Throughout our analysis, we imposed flat priors as follows: $\Omega_{m,0} \in[0,1],\; \xi \in [-0.33, 1], \; w \in [-2, 0]$.

\section{Results}
\label{sec-results}

\begin{figure*}
\centering
\includegraphics[width=0.48\textwidth]{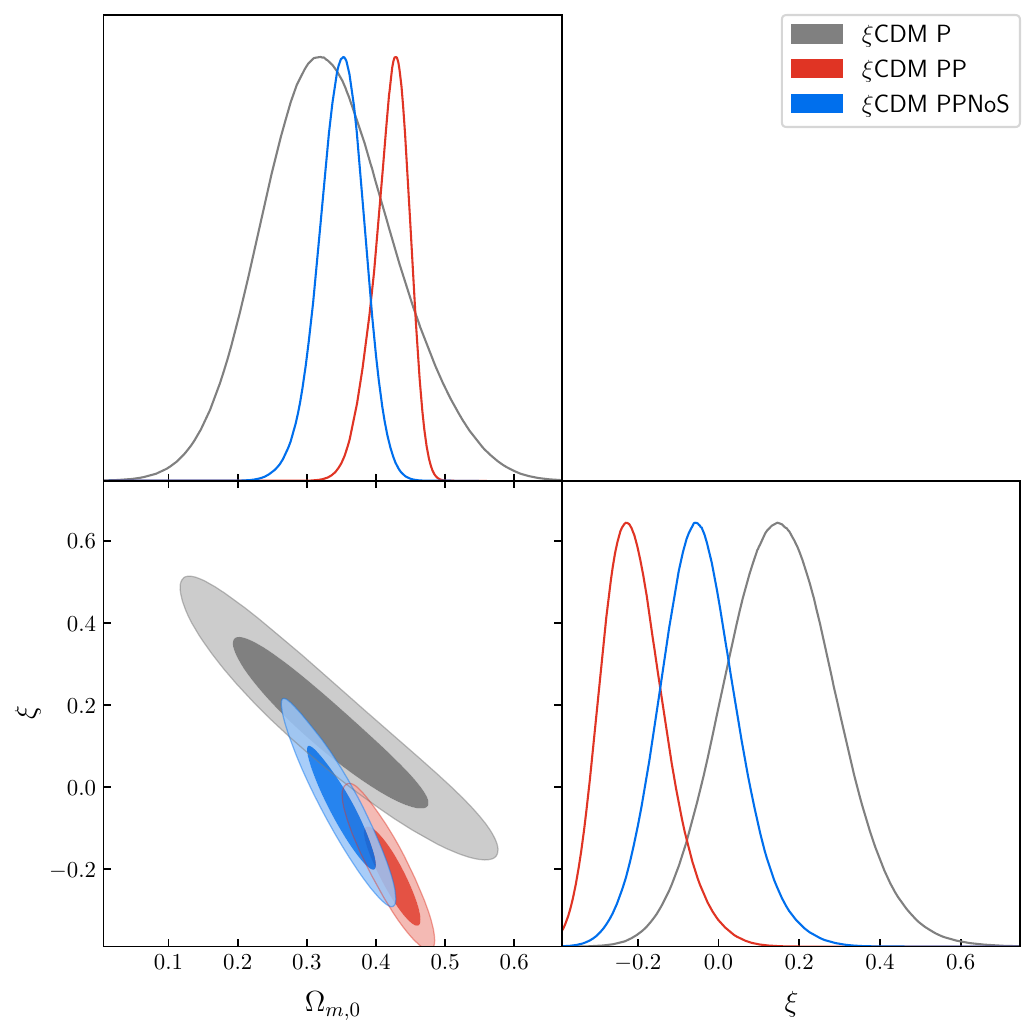} \,\,\,\,\,\,
\includegraphics[width=0.48\textwidth]{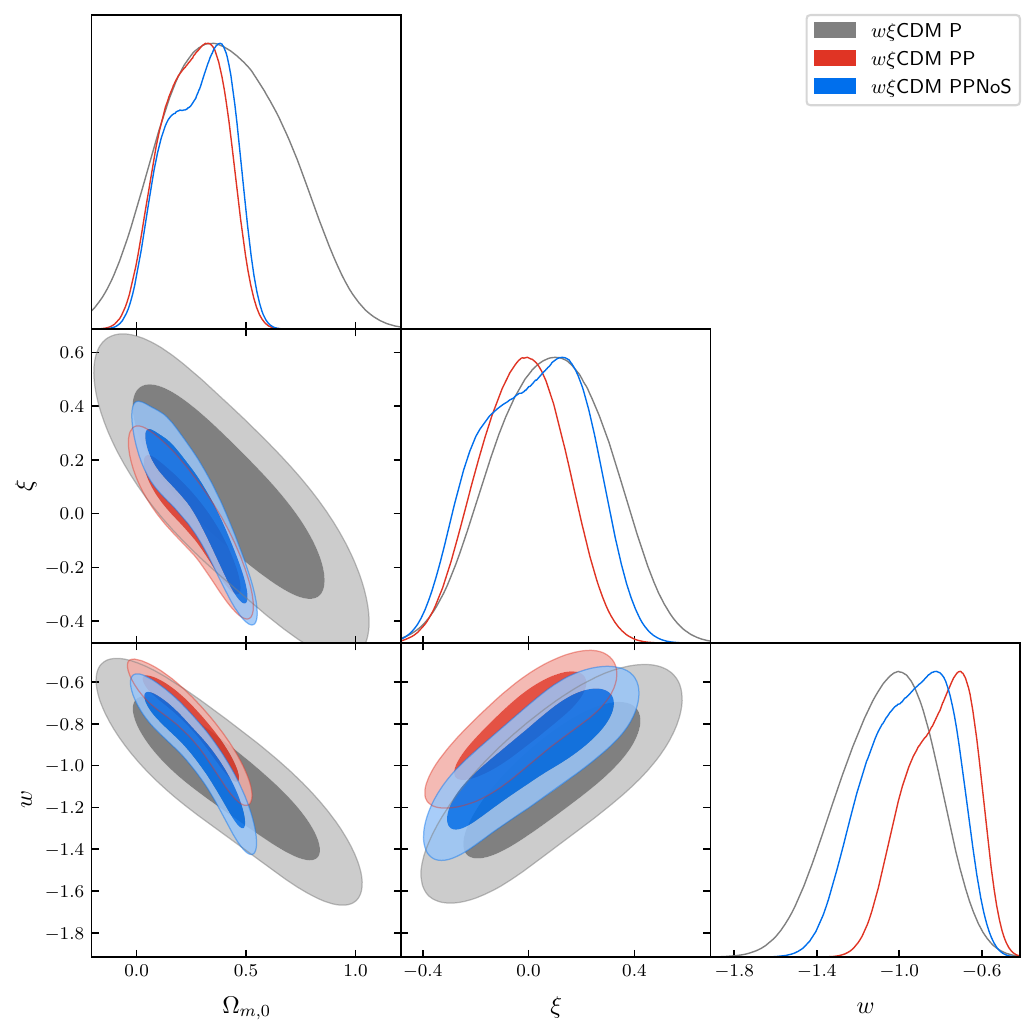} 
\caption{Triangular plots showing the 2D joint and 1D marginalized posterior probability distributions for the free parameters of the $\xi$CDM scenario (left panel) and the $w\xi$CDM scenario (right panel).}
\label{fig:ComparePanth}  
\includegraphics[width=0.49\textwidth]{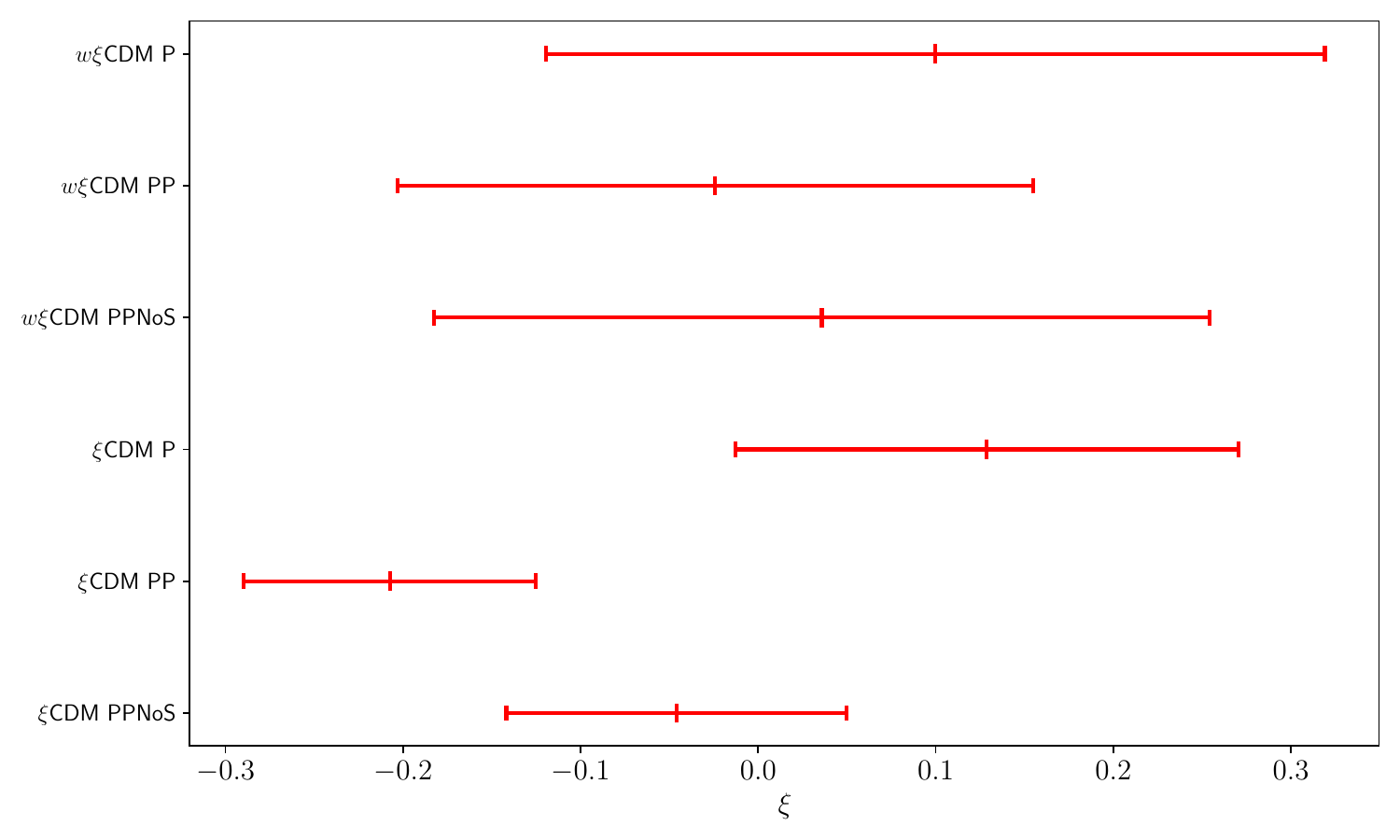}
\includegraphics[width=0.49\textwidth]{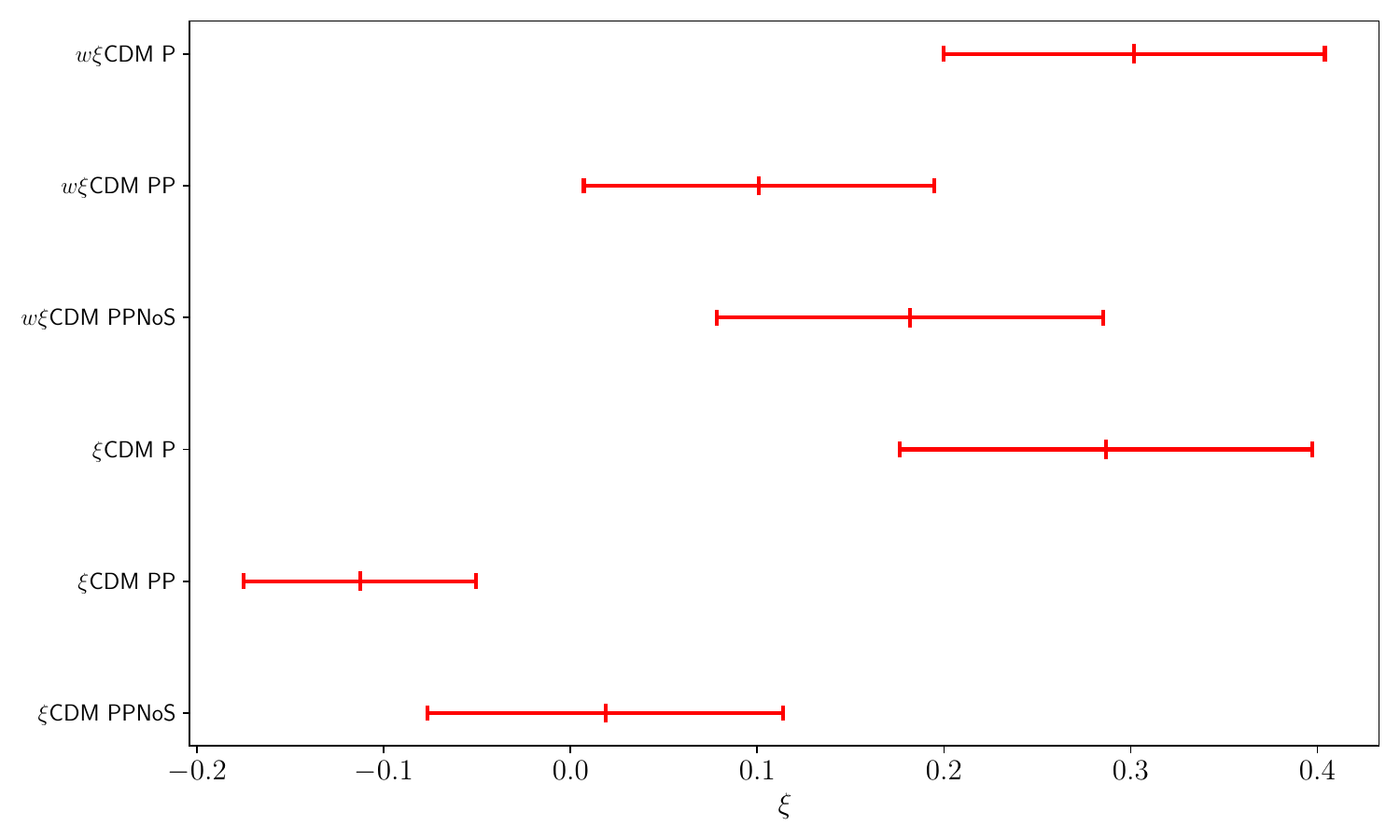}\\ \caption{{Mean values and errors of the parameter $\xi$  presented for the $\xi$CDM and $w\xi$CDM models across the three SNeIa datasets. In most cases, the interaction term agrees with $\xi = 0$ within $1\sigma$. The left panel illustrates the results under a uniform prior on the matter density $\Omega_{\rm m,0} \in [0,1]$, while the right panel assumes a Gaussian prior on $\Omega_{\rm m,0}$ centered around the value obtained from Table 3 of~\cite{Anchordoqui:2021gji}: $\Omega_{m,0}^{CMB}=0.139 \pm 0.095$, derived from an imposed IDE model.}} \label{fig:xi}  
\end{figure*}

In this section, we present the constraints on the interacting scenarios, namely $\xi$CDM and $w\xi$CDM, using the combined datasets including CC, BAO, and various compilations of SNeIa as described in Section~\ref{subsec-data}. After analytically marginalizing the parameters $H_0$, $r_d$, and $M_B$, the free baseline parameters of the $\xi$CDM model 
become $\Omega_{m,0}$ and $\xi$, while for the $w\xi$CDM model, they are $\Omega_{m,0}$, $\xi$, and $w$. Our key results are reported in Table~\ref{Table1} and in Figs.~\ref{fig:ComparePanth} and \ref{fig:xi}.

To infer the constraints on the parameters of these two interacting scenarios, we performed three joint analyzes using three distinct SNeIa samples: PP, PPNoS, and P. This approach allows us to compare the results obtained across three different SNeIa datasets, which can be distinguished by their sample size and systematic astrophysical uncertainties. A similar comparison was conducted in~\cite{Briffa:2023ern}, demonstrating the importance of examining how different SNeIa datasets impact the results when extended beyond $\Lambda$CDM.

As anticipated, we obtained the most precise fit values from the calibrated PantheonPlus and SH0ES (PP) dataset. Here, the inclusion of incredibly precise measurements from the SH0ES collaboration results in very narrow constraints. Conversely, both the PantheonPlus dataset without SH0ES (PPNoS) and the older Pantheon (P) datasets yield broader contours for $\Omega_{m,0}$ and $\xi$ (and $w$), accompanied by slightly less Gaussian convergence. It is important to note that while the marginalization procedure alleviates degeneracies or tensions between certain parameters, this may come at the expense of increased uncertainty in the posterior distribution.

\begin{table}
                \begin{tabular}{cccccccccc}
   \hline
                Model& $\Omega_{m,0}$ & $\xi$  & $w$\\\\
                \hline
       
                \multicolumn{4}{c}{\underline{CC+BAO+PP}} \\\\
    
                        $\xi$CDM & $0.422\pm 0.026$ & $-0.209\pm 0.076$ & $-1$ \\
                        $w\xi$CDM & $0.26\pm 0.16$ & $-0.03\pm 0.17$ & $-0.80\pm 0.18$ \\\\
                        \multicolumn{4}{c}{\underline{CC+BAO+PPNoS}} \\\\
   
                        $\xi$CDM & $0.350\pm 0.031$ & $-0.049\pm 0.096$ & $-1$ \\
                        $w\xi$CDM & $0.28\pm 0.18$ & $0.02\pm 0.24$ & $-0.95\pm 0.23$ \\\\
                        \multicolumn{4}{c}{\underline{CC+BAO+P}} \\\\
                        $\xi$CDM & $0.332\pm 0.089$ & $0.15\pm 0.13$ & $-1$ \\
                        $w\xi$CDM & $0.41\pm 0.30$ & $0.09\pm 0.25$ & $-1.06\pm 0.25$ \\
                        \hline
\end{tabular}
\caption{{Marginalized constraints (mean values with 68\% CL uncertainties) on the free parameters of the interacting scenarios $\xi$CDM and $w\xi$CDM, considering the joint analysis of the CC+BAO+SNeIa dataset for the three SNeIa samples considered.}}
\label{Table1}
\end{table}

We focus on the constraints on the coupling parameter $\xi$ extracted from the two interacting scenarios examined in this study.  Within the $\xi$CDM framework, using the combined dataset CC+BAO+PP, we observe {$\xi = -0.21\pm 0.08$ } at  68\% CL. In this scenario, energy flow from DE to CDM is indicated, resulting in an increase in the CDM energy density throughout the cosmic history. The result at the 95\% CL yields $\xi = -0.21\pm 0.12$, corroborating evidence for $\xi$ and consequently supporting the presence of an interaction
in this context. For the present-day matter density, we obtain $\Omega_{m,0} = {0.42\pm 0.03}$ at the 68\% CL.
However, it is noteworthy that this analysis exhibits a tendency towards a mean negative value of $\xi$ (as discussed below), indicating a higher value of the total matter density compared to the $\Lambda$CDM model. Conversely, when the PPNoS dataset replaces PP in the joint analysis CC+BAO+PPNoS, we find $\xi = {-0.05\pm 0.10}$ at the 68\% CL, indicating complete compatibility with the null hypothesis, that is, the $\Lambda$CDM model. For CC+BAO+PPNoS, we obtain $\Omega_{m,0} = {0.35\pm 0.03} $ at the 68\% CL. Lastly, for the combined dataset employing the Pantheon dataset (CC+BAO+P), we find $\xi = {0.15\pm 0.13} $ at the 68\% CL within the $\xi$CDM framework, indicating the presence of a mild interaction in the dark sector that vanishes at the 95\% CL. The matter density in this scenario closely resembles that of the PPNoS dataset, leading to $\Omega_{m,0}=0.33 \pm 0.09 $ at the 68\% CL.

Regarding the estimations of the Hubble constant, as highlighted in~\cite{Dhawan:2020xmp,Brout:2022vxf}, the local $H_0$ constraint derived from the Cepheid distance ladder remains insensitive to models beyond the $\Lambda$CDM cosmology. However,  the strong correlation between $\xi$ and $H_0$ is well established (as discussed in~\cite{Zhai:2023yny} and references therein). Consequently, during the marginalization process over $H_0$ when conducting joint analyses with PP samples--- which incorporate Cepheid distance measurements--- it is expected that the statistical information regarding the correlation with $H_0$ will be preserved, thereby maintaining a tendency for $\xi < 0$ at the 68\% CL.
Conversely, the analysis using the P samples does not extend to cover the low redshifts of the primary distance indicators. Therefore, in this analysis, a tendency for $\xi < 0$ is not expected. Interestingly, a notable trend towards $\xi > 0$ is observed with P samples, suggesting that SNeIa samples lacking primary distance indicators at very low redshifts may indicate a tendency for the coupling parameter to change sign. However, the analysis with PPNoS, which represents an updated version of P samples with an increased sample size, demonstrates complete compatibility with $\xi = 0$.

Now, shifting our focus to the $w\xi$CDM scenario, we do not find any evidence supporting $\xi \neq 0$ in the context of the joint analyses CC+BAO+PP, CC+BAO+PPNoS, and CC+BAO+P.
It is worth noting that when employing a narrower prior on $\Omega_{m,0}\in [0.2,0.4]$, as detailed in Appendix A, we find some tentative evidence supporting an interaction in the dark sector at more than 95\% confidence level (CL) for CC+BAO+P ($\xi = 0.18 \pm 0.17$ at 95\% CL).
Furthermore, focusing on the dark energy equation of state, $w$, we observe that for the joint analyzes with the PPNoS and P datasets, the constraints are fully compatible with $w = -1$. However, in the case of the joint analysis with the PP dataset, we find indications of a quintessence-type behavior at the 68\% CL.

\begin{table}
\centering
                \begin{tabular}{cccccccccc}
   \hline
   Model & $\Omega_{m,0}$ & $\xi$ & $w$ \\
   \hline
                        \multicolumn{4}{c}{\underline{CC+BAO+PP}} \\\\
                        $\xi$CDM & $0.389\pm 0.018$ & $-0.113\pm 0.062$ & $-1$ \\
                        $w\xi$CDM & $0.144\pm 0.082$ & $0.101\pm 0.094$ & $-0.680\pm 0.092$ \\\\

                        \multicolumn{4}{c}{\underline{CC+BAO+PPNoS}} \\
                        $\xi$CDM & $0.325\pm 0.031$ & $0.019\pm 0.095$ & $-1$ \\
                        $w\xi$CDM & $0.150\pm 0.090$ & $0.18\pm 0.10$ & $-0.774\pm 0.092$ \\\\

                        \multicolumn{4}{c}{\underline{CC+BAO+P}} \\
                        $\xi$CDM & $0.243\pm 0.061$ & $0.29\pm 0.11$ & $-1$ \\
                        $w\xi$CDM & $0.130\pm 0.096$ & $0.30\pm 0.10$ & $-0.849\pm 0.080$ \\

   \hline
\end{tabular}
\caption{{Similar to Table I, but in this case, we are assuming a Gaussian prior on the matter density $\Omega_{\rm m,0}$ centered around the CMB value from  Table 3 of~\cite{Anchordoqui:2021gji}: $\Omega_{m,0}^{CMB}=0.139 \pm 0.095$ at 68\% CL.
}}
\label{Table2}
\end{table}

It is noteworthy that in the $w$CDM model, a tendency towards $w > -1$ was highlighted by~\cite{Brout:2022vxf} based solely on the PP analysis.
With the addition of CC data, which can also lead to values of $w$ tending towards $w > -1$ in the $w$CDM model~\cite{escamilla2023state}, we observe that this preference for $w > -1$ persists in our analysis within the framework of the $w\xi$CDM model as well.

To examine the impact of the prior on the results, we performed a joint analysis using the CC+BAO+SNeIa dataset for the three SNeIa samples considered, assuming a Gaussian prior on the matter density $\Omega_{\rm m,0}$ centered around the CMB point from Table 3 of~\cite{Anchordoqui:2021gji}; that is, $\Omega_{m,0}^{CMB}=0.139 \pm 0.095$, which was derived under the assumption of an IDE model.
The results (presented in Table~\ref{Table2}) reveal that for the CMB prior, indications of IDE at the 68\% CL are observed for the $\xi$CDM model with the CC+BAO+PP dataset and for the CC+BAO+P dataset. Regarding the $w\xi$CDM model, evidence for IDE is obtained across all datasets. At the 95\% CL, evidence for IDE is observed for $\xi$CDM with the CC+BAO+P and CC+BAO+PP datasets, and for $w\xi$CDM with the CC+BAO+P dataset.

Finally, in Fig. \ref{fig:xi}, we compare the mean values and errors for $\xi$ at the 68\% CL under two different priors on $\Omega_{m,0}$: the standard uniform flat prior and a Gaussian prior corresponding to the CMB prior discussed above. It is evident that, while changing the prior reduces the errors, the final mean values remain similar.

An important point to note is the integration of $E(z)$ for negative values of $\xi$, which may lead to numerical instabilities in nontrivial regions of the parameter space. For this reason, our prior on $\xi$ is not symmetric with respect to zero. We chose a left boundary for $\xi$ to avoid  numerical singularities arising from $E(z)$ becoming imaginary. Additionally, we conducted further tests using a normalized $\chi^2$ and a Cholesky decomposition of the covariance matrix for the PPnoS dataset, which provides an alternative method of computing the inverse matrix. However, neither method resulted in improved convergence or smaller errors.

                \begin{table}
        \begin{center}
                \begin{tabular}{ccccccccccccc}
                        \hline
                        Model &~~~~~~~~ $\Delta$AIC &~~~~~~~~ $\Delta$BIC &~~~~~~~~ $\ln$(BF) \\
            \hline 
             \multicolumn{4}{c}{\underline{CC+BAO+PP}} \\\\
                        $\Lambda$CDM &  $-$ &  $-$ & $-$ \\
                        $\xi$CDM & 0.74 & $-4.72$ & $-1.60$ \\
                        $w\xi$CDM & $-1.34$ & $-12.26$ & $-0.16$ \\\\
                        \multicolumn{4}{c}{\underline{CC+BAO+PPNoS}} \\\\
                        $\Lambda$CDM &  $-$ &  $-$ & $-$ \\
                        $\xi$CDM & $ -2.21$ & $-7.60$ & 0.97 \\
                        $w\xi$CDM & $-4.29$ & $-15.06$ & 2.26 \\\\
                        \multicolumn{4}{c}{\underline{CC+BAO+P}} \\\\
                        $\Lambda$CDM &  $-$ &  $-$ &  $-$ \\
                        $\xi$CDM & $-1.99$ & $-4.24$ & 0.79 \\
                        $w\xi$CDM & $-3.92$ & $-8.41$ & 1.92 \\
                        \hline
                \end{tabular}
        \end{center}
        \caption{{Summary of the information criteria for the $\xi$CDM and $w\xi$CDM models compared to the flat $\Lambda$CDM model, utilizing the combined dataset CC+BAO+SNeIa with consideration of three distinct samples of SNeIa. The AIC indicates a preference for $\Lambda$CDM, albeit weaker when Cepheids are included. Conversely, the BIC strongly favors $\Lambda$CDM. Additionally, the BF demonstrates a weak preference for the IDE models in the combined analysis CC+BAO+PP. 
 }}
\label{Table3}
\end{table} 

To compare the different models using statistical measurements assuming the defined datasets, we employed the Akaike information criterion (AIC), the Bayesian information criterion (BIC), and the Bayes factor (BF)~\cite{Liddle:2007fy,Staicova:2021ntm}. The AIC criterion is defined as:
        \begin{equation}
        \text{AIC}=-2\ln(\mathcal{L}_{\text{max}})+2k+
        \frac{2k(k+1)}{N_{\rm tot}-k-1}\,,
        \end{equation}
where $\mathcal{L}{\text{max}}$ represents the maximum likelihood of the data under consideration, $N_{\rm tot}$ is the total number of data points, and $k$ is the number of parameters. The BIC criterion is defined as
\begin{equation}
\text{BIC} = -2\ln(\mathcal{L}_{\text{max}})+k \,{\rm log}(N_{\text{tot}})\,.
\end{equation} 
From these definitions, we calculated $\Delta \text{IC}{\text{model}}=\text{IC}_{\Lambda\text{CDM}}-\text{IC}_{\text{model}}$, where our base model is the flat $\Lambda$CDM. The model with the minimal AIC is considered the best~\cite{Jeffreys:1939xee}, with a positive $\Delta$IC giving preference to the IDE model, and a negative $\Delta$IC favoring $\Lambda$CDM, with $|\Delta\text{IC}|\geq 2$ signifying a possible tension. The logarithmic BF is defined as: 
\begin{equation}
\ln\left(B_{0j}\right)= \ln\left[\frac{p(d|M_0)}{p(d|M_j)}\right],
\end{equation}
where $p(d|M_j)$ is the Bayesian evidence for model $M_j$. The evidence is calculated numerically by Polychord. In Table~\ref{Table3}, we set $M_0=M_{\Lambda CDM}$, which we compare to the IDE models. According to the revised Jeffrey's scale~\cite{Jeffreys:1939xee}, $|\ln(B_{0j})| <1$ is inconclusive for any of the models, negative values support the IDE model, and positive values favor the $\Lambda$CDM model.

Our results are summarized in Table \ref{Table3}. For all three datasets, we compared the $\Lambda$CDM model with the IDE models we consider. We observe that the AIC and BIC criteria strongly favor $\Lambda$CDM. The only weak support is for $\xi$CDM PP. The BF mostly favors $\Lambda$CDM, with the notable exception of $\xi$CDM PP. In this case, we also observe evidence for IDE from $\xi$ at 95\% CL. The other case in which there is a preference for $\xi$CDM at 68\% CL, which is Pantheon, shows no statistical preference for $\xi$CDM, but also inconclusive evidence for $\Lambda$CDM.
Thus, from a statistical standpoint, for most measures, no notable preference  is observed in any of the present interacting scenarios. 
However, the results are sensitive to the underlying interacting model.

\section{Discussion and future prospects}
\label{sec-conclusions}

The interaction between DM and DE is a well-known cosmological scenario that has garnered enormous attention in the community. As explored in the literature, IDE models play an effective role in reconciling the $H_0$ tension, and this reconciliation is related to the underlying interacting model and its parameters. Additionally, parameters such as the sound horizon, $r_d$, and the absolute magnitude, $M_B$, are also related to the $H_0$ tension. Consequently, all three parameters --- $H_0$, $r_d$, and $M_B$ --- are dependent on the interacting parameters.
In this work, we investigated an interacting model using a heuristic approach that allows us to examine the intrinsic nature of the model parameters without directly linking them to $H_0$, $r_d$, or $M_B$. We employed the marginalization method to remove the variables $H_0$, $M_B$, and $r_d$, with the aim being to examine the intrinsic nature of the interaction between the dark components. We used a transversal BAO dataset, cosmic chronometers (CC) whilst accounting for covariance, and different compilations of SNeIa datasets ---PantheonPlus and SH0ES, PantheonPlus and SH0ES without SH0ES prior (uncalibrated and calibrated), and the old Pantheon--- for the purpose of comparing results.

Considering a well-known interaction function of the form $Q = 3 H \xi \rho_{\rm DE}$, we investigated two distinct interacting scenarios, labeled $\xi$CDM and $w\xi$CDM. We constrained both scenarios using the combined datasets CC+BAO+PP, CC+BAO+PPNoS, and CC+BAO+P, and the results are summarized in Table~\ref{Table1} and Fig.~\ref{fig:ComparePanth}. Our results show that, for the uncalibrated PantheonPlus and SH0ES (i.e., PPNoS), when combined with CC and BAO, we do not find any evidence of $\xi \neq 0$ in either the $\xi$CDM ($\xi = {-0.05\pm 0.1}$ at 68\% CL, CC+BAO+PPNoS) or $w\xi$CDM ($\xi = {0.02\pm 0.24}$ at 68\% CL, CC+BAO+PPNoS) scenario.
In the calibrated case (PP), evidence of $\xi \neq 0$ is found in $\xi$CDM at more than 68\% CL ($\xi = -0.21\pm 0.08$ for CC+BAO+PP), indicating a flow of energy from DE to DM. This is further confirmed by the high value of the matter density parameter ($\Omega_{m,0} = 0.42\pm 0.03$ at 68\% CL, CC+BAO+PP), which remains at 95\% CL. However, for the $w\xi$CDM scenario, we do not find any such statistical evidence, as reflected by the coupling parameter $\xi = -0.02 \pm 0.17$ at 68\% CL (CC+BAO+PP).
On the other hand, when the Pantheon dataset is used (i.e., for the combined dataset CC+BAO+P), we obtain $\xi = 0.15\pm 0.13$ at 68\% CL for $\xi$CDM and $\xi = 0.09\pm 0.25$ at 68\% CL for $w\xi$CDM, indicating a preference for an interaction between DE and CDM for $\xi$CDM. However, within the 95\% CL, the evidence for $\xi \neq 0$ diminishes, eventually recovering $\Lambda$CDM and $w$CDM models, respectively.
Finally, we observe that $w$ is closest to $\Lambda$CDM for the uncalibrated PantheonPlus and SH0ES dataset.

The marginalization procedure yields interesting new results that exhibit a relatively strong dependence on whether the SNeIa dataset is calibrated with the Cepheids or not. Additionally, we observe significant differences between the Pantheon and PantheonPlus datasets in terms of the uncertainties, indicating that PP and PPNoS are  more suitable for this kind of study. While IDE demonstrates exciting potential to alleviate the Hubble tension, further studies on the choice of datasets and parameter space are needed to confirm its contribution. Notably, in this work, we exclude the CMB contribution and only use the transversal BAO dataset, which, while more suitable due to its independence of fiducial cosmology, leads to larger errors compared to the newest mixed angular and radial BAO datasets.
In summary, our results imply that
the marginalization methodology adopted to examine this particular interacting model could provide new insights if 
applied to other promising interacting models and with more datasets.

\section*{Acknowledgments}
D.B. thanks the Carl-Wilhelm Fueck Stiftung and the Margarethe und Herbert Puschmann Stiftung. S.P. acknowledges the financial support from  the Department of Science and Technology (DST), Govt. of India under the Scheme ``Fund for Improvement of S\&T Infrastructure (FIST)'' [File No. SR/FST/MS-I/2019/41]. D.S. acknowledges the support of Bulgarian National Science Fund No. KP-06-N58/5. E.D.V acknowledges support from the Royal Society through a Royal Society Dorothy Hodgkin Research Fellowship. R.C.N thanks the financial support from the Conselho Nacional de Desenvolvimento Cient\'{i}fico e Tecnologico (CNPq, National Council for Scientific and Technological Development) under the project No. 304306/2022-3, and the Fundação de Amparo à pesquisa do Estado do RS (FAPERGS, Research Support Foundation of the State of RS) for partial financial support under the project No. 23/2551-0000848-3. This article is based upon work from the COST Action CA21136 "Addressing observational tensions in cosmology with systematics and fundamental physics" (CosmoVerse), supported by COST (European Cooperation in Science and Technology).

\bibliographystyle{aa}
%
\bibliography{ref.bib}

\begin{appendix}

\section*{Appendix A - Assuming smaller priors on the matter density}
\label{sec-appendix}

\begin{figure}[t!]
\centering 
\includegraphics[width=0.45\textwidth]{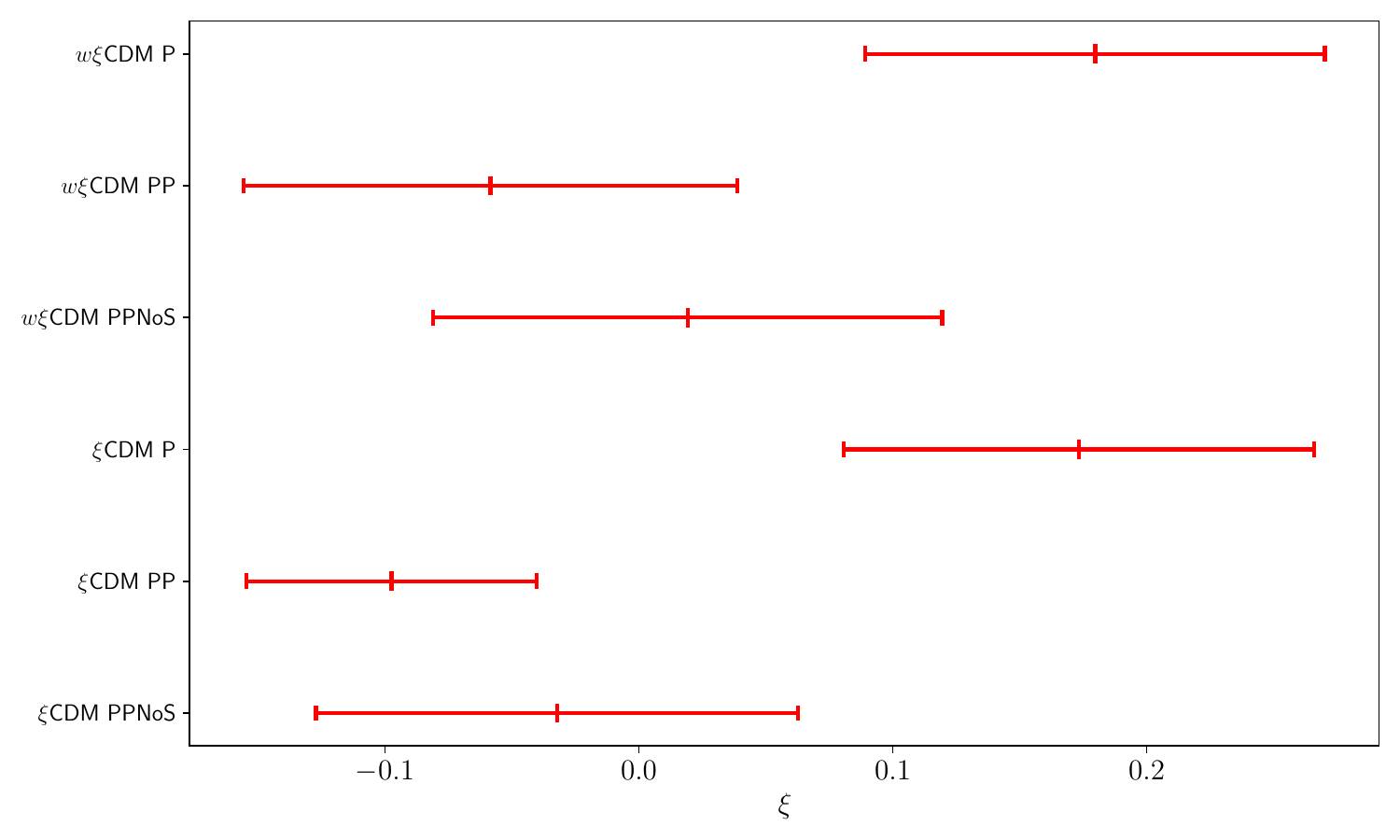}
\caption{{Mean values and errors of the parameter $\xi$ for the $\xi$CDM and $w\xi$CDM models for the three SNeIa datasets with a uniform prior on the matter density parameter  $\Omega_{\rm m,0} \in [0.2,0.4]$. In most cases, the interaction agrees with $\xi = 0$ and this is consistent within $1\sigma$ uncertainty. }}
\label{fig:xi_sm}  
\end{figure}

For completeness, here we report  
the marginalized constraints  on the $\xi$CDM and $w\xi$CDM scenarios considering a smaller prior on $\Omega_{m,0}\in [0.2,0.4]$. Table~\ref{Table4} and Fig.~\ref{fig:xi_sm} summarize the results on these two interacting scenarios. 
In this case, we observe evidence of interaction within $\xi$CDM  for  CC+BAO+PP and CC+BAO+P datasets, while for $w\xi$CDM, evidence is found only for CC+BAO+P. Within the 95\% CL, the results that remain are $\xi = 0.17 \pm 0.15$ for $\xi$CDM and $\xi= 0.18 \pm 0.17$ for $w\xi$CDM, both for the CC+BAO+P dataset. For the rest, the results are consistent with no interaction within 95\% CL.

\begin{table}
        \centering
        \begin{tabular}{c c c c c c c c c c c}
                \hline
                Model& $\Omega_{m,0}$ & $\xi$  & $w$\\
                \hline
     
                \multicolumn{4}{c}{\underline{CC+BAO+PP}} \\\\
        
     $\xi$CDM  & $0.381\pm 0.015$ & $-0.098\pm 0.061$ & $-1$\\

     $w\xi$CDM & $0.309\pm 0.065$ & $-0.058 \pm 0.097$ & $-0.834\pm 0.087$ \\

     \\
        
     \multicolumn{4}{c}{\underline{CC+BAO+PPNoS}} \\\\

     $\xi$CDM & $0.343\pm 0.028$ & $-0.032\pm 0.095$ & $-1$ \\

        $w\xi$CDM & $0.302\pm 0.069$ & $0.02\pm 0.10$ & $-0.95\pm 0.11$ \\
        
        \\

        \multicolumn{4}{c}{\underline{CC+BAO+P}} \\\\

     $\xi$CDM  & $0.316 \pm 0.062$ & $0.173\pm 0.093$ & $-1$ \\

     $w\xi$CDM  & $0.290\pm 0.061$ & $0.180\pm 0.090$ & $-0.971\pm 0.079$ \\

                \hline
        \end{tabular}
 \caption{{Marginalized constraints (mean values with their 68\% CL uncertainties) on the free parameters of the interacting scenarios $\xi$CDM and $w\xi$CDM considering the joint analysis of the CC+BAO+SN dataset for the three SN samples we consider, with the prior $\Omega_{\rm m,0} \in [0.2,0.4]$.} }
 \label{Table4}
\end{table}
%


\end{appendix}


\end{document}